\pdfoutput=1

\documentclass[11pt]{article}

\usepackage{acl}
\usepackage{mathtools}

\usepackage{times}
\usepackage{latexsym}

\usepackage[T1]{fontenc}

\usepackage[utf8]{inputenc}
\usepackage{array,graphicx,dcolumn,multirow,abstract,hanging,fancyhdr,float}
\usepackage{natbib}

\usepackage{microtype}
\usepackage{xcolor}
\usepackage{hyperref}
\usepackage{booktabs}

\title{Designing weighted and multiplex networks for deep learning user geolocation in Twitter}

\author{Federico M. Funes \\
  Facultad de Ingeniería (UBA) \\
  \texttt{ffunes@fi.uba.ar} \\\And
  José Ignacio Alvarez-Hamelin \\
  Facultad de Ingeniería (UBA) \\
  INTECIN (UBA-CONICET) \\
  \texttt{ihameli@cnet.fi.uba.ar} \\\AND
  Mariano G. Beiró \\
  Facultad de Ingeniería (UBA) \\
  INTECIN (UBA-CONICET) \\
  \texttt{mbeiro@fi.uba.ar} \\}
  
\begin{document}
\maketitle
\begin{abstract}

Predicting the geographical location of users of social media like Twitter has found several applications in health surveillance, emergency monitoring, content personalization, and social studies in general.
In this work we contribute to the research in this area by designing and evaluating new methods based on the literature of weighted multigraphs combined with state-of-the-art deep learning techniques. The explored methods depart from a similar underlying structure (that of an extended mention and/or follower network) but use different information processing strategies, e.g., information diffusion through transductive and inductive algorithms --RGCNs and GraphSAGE, respectively-- and node embeddings with Node2vec+.
These graphs are then combined with attention mechanisms to incorporate the users' text view into the models.
We assess the performance of each of these methods and compare them to baseline models in the publicly available Twitter-US dataset; we also make a new dataset available based on a large Twitter capture in Latin America.
Finally, our work discusses the limitations and validity of the comparisons among methods in the context of different label definitions and metrics.
\end{abstract}

\section{Introduction}

User geolocation in social media platforms has proven a useful resource for the development of tools for public health surveillance (e.g., monitoring of general ailments and symptoms \citep{paul2011you}; forecasting of zika virus incidence \citep{mcgough2017forecasting}) and of emergency monitoring systems (e.g., earthquakes \citep{sakaki2010earthquake}; forest fires \citep{de2009omg}; floods \citep{jongman2015early}). However, many social media such as Twitter do not provide wide access to their users' location information: i.e., the location field in the Twitter user profile is unstructured and noisy, and thus difficult to map to a real location. It has also been assessed that only $\sim1\%$ of the tweets are geotagged~\citep{cheng2010you}, making it challenging to automatically link the large flow of real-time information available through the platform to specific places.

These limitations motivate the study of location prediction in Twitter based on the information that is available on the platform, such as the content published by the users, the hashtags they use, and the users they mention, reply to, or follow. This task can be enriched by combining techniques from a variety of areas, such as Natural Language Processing, social networks and graph mining.

The literature on user geolocation in Twitter has developed several graph-based and content-based deep learning models in the last years, while it also proposed the usage of specific metrics as \textit{Acc@100}, and alternative label construction techniques as the usage of \textit{k-d} trees. Many works also explored the usage of probabilistic classifiers and information-theoretic or statistic measures for detecting Location Indicative Words (LIW's), and the remotion of celebrities from the contact graphs. We make a brief historical account of these advances in Section~\ref{sec_rel_work}.

\section{Data collection and preprocessing}

In this work we introduce a dataset collected between January 1, 2019 and December 10, 2019 in Argentina using the Twitter API, in the context of a social study~\cite{reyero2021evolution}. The users' set was defined as those Twitter accounts who were following at least one of the candidates for the Argentinian presidential elections during 2019, which made for a total of 900 million tweets captured, belonging to 2 million active users. For each user, all her tweets were collected, unless she had posted more than 3,000 tweets between two consecutive polls (around 5 days) --something which happened with extremely low frequency-- in which case her tweets were automatically capped by the API.

From these 900 million tweets, around 1 million were precisely geolocalized by a set of coordinates in the metadata; this statistic is consistent with previous findings (e.g.,~\citep{cheng2010you}). Moreover, 14 million tweets had metadata that geolocalized them inside a bounding box and mentioned a specific city chosen by the user when tweeting. We unified and validated the latter by combining the information with the Geonames gazzetteer (\url{https://www.geonames.org/}), assigning a city in Geonames and a set of precise coordinates to each geolocated tweet. In this process, around 2 million tweets had to be discarded.

For each user with at least one geolocated tweet, we defined her ground-truth location as the city in which most of her geolocated tweets were placed. At this point, we built the two labelled datasets described in Table~\ref{tab:our_datasets}. The tweet ids composing each of these datasets are publicly available at GitHub\footnote{\href{https://github.com/fedefunes96/twitter-location-data.git}{https://github.com/fedefunes96/twitter-location-data.git}} for hydration. Compared to previous datasets in the literature, our datasets offer a higher average of tweets per user (around $1,000$), thus providing a large volume of content data for training.

Our datasets \texttt{Twitter-ARG-Exact} and \texttt{Twitter-ARG-BBox} cover $95$ and $229$ cities accross $14$ and $22$ different countries respectively (with at least $100$ users in each city), but most of the users ($76.9\%$ and $72.6\%$ respectively) lie inside Argentina.

\begin{table*}[]
    \centering
    \begin{tabular}{l|lll}
    \toprule
        Dataset & Users & Geolocated tweets & Total tweets \\
    \midrule
     \texttt{Twitter-ARG-Exact} & $37,146$ & $625,180$ & $27,574,343$ \\
        \texttt{Twitter-ARG-BBox} & $141,209$ &  $9,298,954$ & $124,192,146$ \\
    \bottomrule
    \end{tabular}
    \caption{Basic statistics for the datasets introduced in this work.}
    \label{tab:our_datasets}
\end{table*}

\subsection{Profile location analysis}

We analyzed the veracity of the location informed by the users in their profile. ~\citep{cheng2010you} had found that only $26\%$ of the users mention a city in their profile location field. We applied a tokenizer to the location field and tried to match the words to a city and/or country name in GeoNames. We found a match with one or more GeoNames cities for $63\%$ of our users who had some profile location information available. Interestingly, for those with only one match ($40.9\%$ of the total users), it turned out that $52\%$ of them were coincident with the ground-truth location obtained from their georeferenced tweets. All in one, this implies that only $20\%$ of our users have an unambiguous city in their location profile that matches their geolocated tweets.

\begin{figure}[t]
\centering
\includegraphics[width=7.5cm]{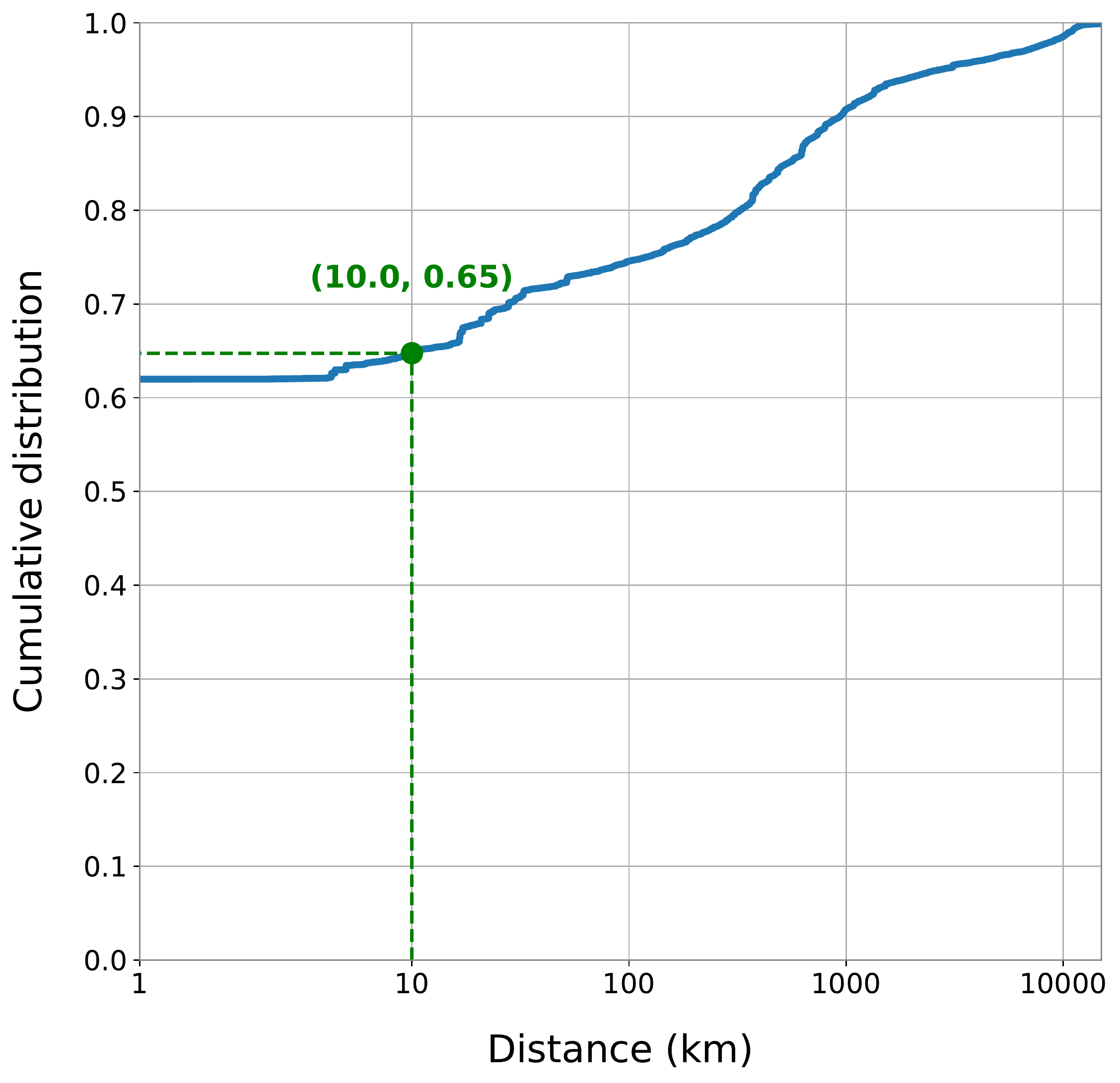}
\caption{\label{distance_distribution}Cumulative distribution of the distance between the city in the user profile and the ground-truth city given by the majority of the user's geotagged tweets. These results are in agreement with~\citep{cuevas2014understanding}.}
\end{figure}

Figure~\ref{distance_distribution} shows the distribution of distances between the profile city and the ground-truth one extracted from the geolocated tweets, for those users whose profile city is unambiguous. We observe that $65\%$ of these users have an error of less than 10km between their profile location and their ground-truth one. However, $25\%$ of the users have an error larger than $161\text{km}$, which is the typical relaxed accuracy bound in the location prediction literature, and thus we consider that the profile location is not accurate enough for building ground-truth city labels.

\subsection{Graphs' construction}

Exploiting the mentions contained in the users' tweets, and the users' following relationships extracted through the Twitter API, we built user networks that can help predict their locations.

Frequently, users in our dataset will follow or mention users which are not part of it. Although we could discard these connections, they might be relevant for inferring  the user's location: e.g., if two users follow many people in common, their probability of being geographically close is larger than for a pair of random users.

Given the set of users in our dataset, $U$, and the extended set of users $E$ obtained by adding users mentioned by those in our dataset, we denote $n=|U|$ and $m=|E|$, and we consider the mention matrix $M \in \mathbf{R}^{n\times n}$ which represents a weighted directed graph with no loops, whose nodes are those users in $U$. The elements $(m_{ij})$ in this matrix point out the number of times that user $i$ mentioned user $j$ in her tweets. 

Previous works in the literature obtain a \textit{collapsed} mention matrix that adds co-mentions to the mention network, by computing $(M + I)\cdot (M + I)^T$ (see,  e.g., ~\cite{rahimi-etal-2015-twitter,P18-1187}).



In this work we propose an alternative solution that combines the connections in the mention graph $M$ with mention paths that traverse users outside our user set; this procedure gives a weighted graph as result. We define our \textit{extended mention network} as $Y=(M + M^T) + (X_M\cdot X_M^T - diag(X_M\cdot X_M^T))$. Here, $X_M\in \mathbf{1}^{n\times (m-n)}$ represents the existence of mentions from users in $U$ to users not in $U$, and thus corresponds to a bipartite graph. The left term in the formula represents the symmetrized mention matrix (where each element $a_{ij}$ represents the number of times that users $i$ and $j$ mention one another), while the right term represents the number of users external to our dataset that both $i$ and $j$ mention. Thus, the latter extends the mention network by adding co-mentions going through external users. Following~\cite{rahimi-etal-2015-twitter} we remove mentions to popular users before computing our extended network.

\begin{figure}[t]
\centering
\includegraphics[height=4cm]{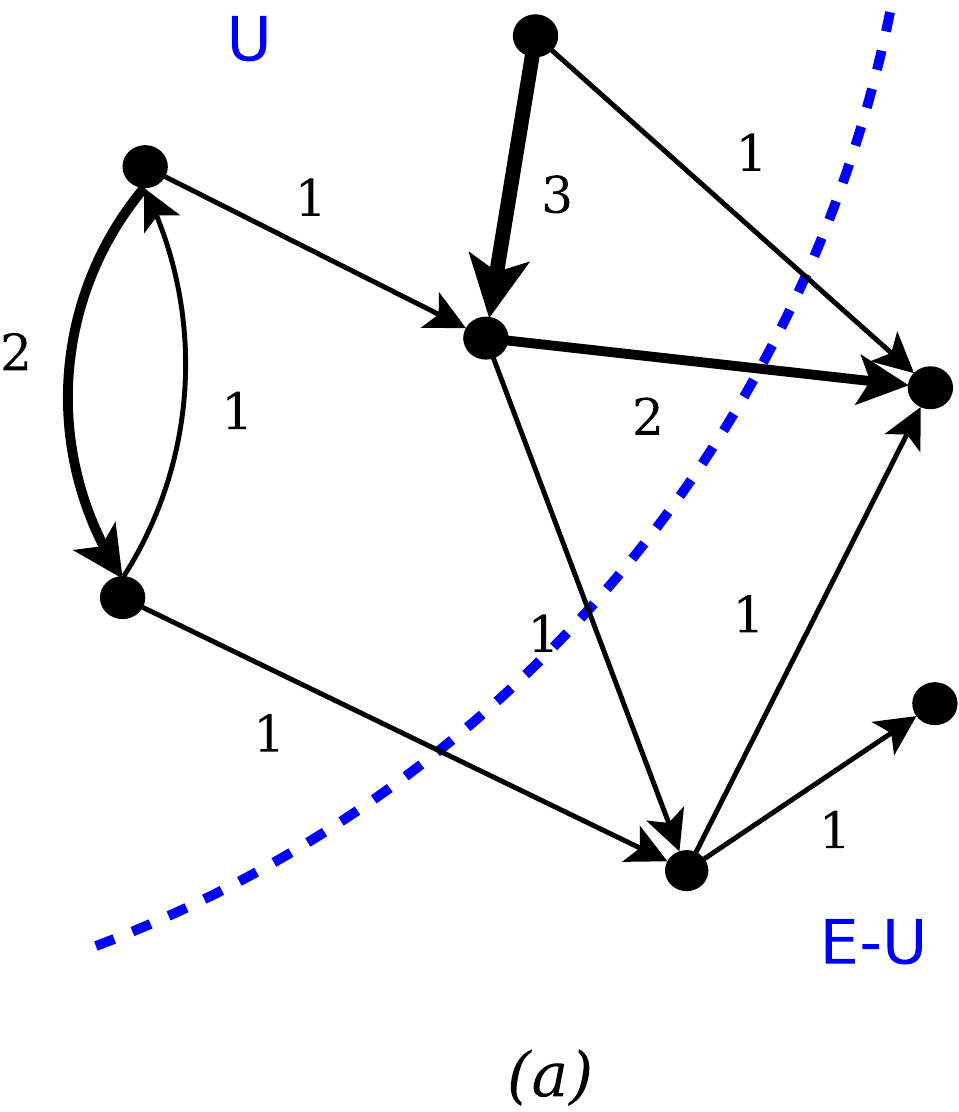}
\hspace{0.4cm}
\includegraphics[height=4cm]{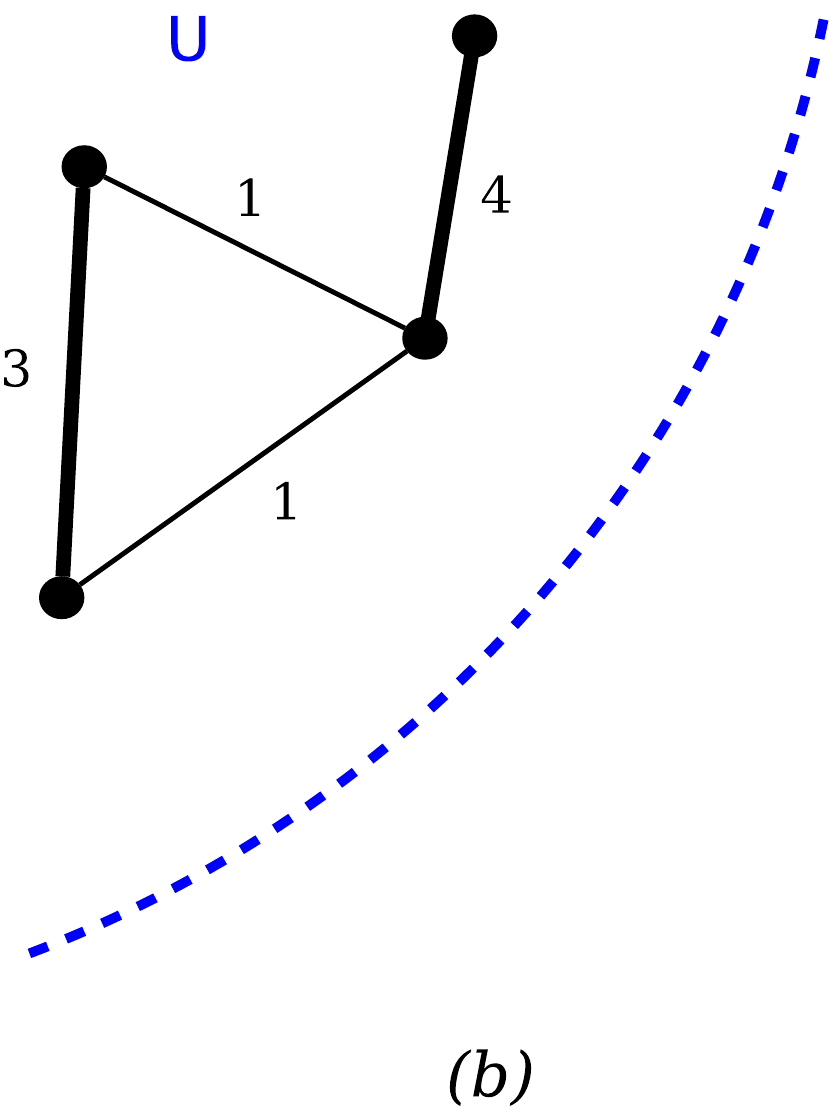}
\caption{\label{fig:mentions_graph}Construction of the extended mention graph. \textit{(a)} Original mention graph, including mentions to users not in $U$. \textit{(b)} Extended mention graph for the set of users $U$ in our dataset.}
\end{figure}

The advantages of this extended network with respect to the previous collapsed networks' definitions are two-fold: firstly, it can simultaneously weight mentions and co-mentions in the results --thus profiting of all the mention information that we collect in tweets-- and secondly, that it reduces the complexity of the obtained network --in~\cite{rahimi-etal-2015-twitter} and~\cite{P18-1187}, for example, $n$ nodes mentioning a central node in the training/test set generate $n\cdot(n-1)$ edges in the collapsed network--. We illustrate our procedure in Figure~\ref{fig:mentions_graph}.

When the follower network is available --as is the case in our dataset-- we also extend it by including connections to users outside our dataset. 
We define our \textit{extended follower network} as $Z=F + (X_F\cdot X_F^T - diag(X_F\cdot X_F^T))$. Here, $X_F\in \mathbf{1}^{n\times (m-n)}$ represents follower relationships from users in $U$ to users not in $U$. Put into words, this computation extends the follower network by adding co-follower relationships to external users (i.e., counting the number of users outside our dataset that are co-followed by a pair of users $i$ and $j$ in our dataset). We treat the resulting network as undirected and weighted, and we also remove popular nodes from the original data before computing $Z$.

In one of our models, \texttt{RGCN-EXT}, both the extended mention graph and the extended follower graph are combined into a multiplex graph~\cite{kivela2014multilayer}. In this type of graph, the set of nodes is the same across all layer, but each layer represents a different relationship type; thus, the network can couple the information diffusion paths provided by each layer. This could be further generalized to allow the flow of information across layers as in general multilayer graphs, or to decompose layers into sublayers (e.g., dividing the mention layer into pure mentions and external comentions).


\subsection{Labels construction}

In most cases, ground-truth location labels lie on a continuous space given by the precise latitude and longitude coordinates --e.g., in our \texttt{Twitter-ARG-Exact} dataset--. In order to discretize these labels,~\citep{roller2012supervised} proposed the usage of \textit{k-d} trees~\cite{bentley1975multidimensional}. This method builds equally-sized partitions of space which can improve the classifiers' learning task.

The usage of \textit{k-d} trees also has some pitfalls: if many users share the same set of coordinates (i.e., due to precision limitations), it becomes hard to obtain similarly sized fine-grained partitions; also, in the presence of very dense metropolitan areas, they might be split into several \textit{k-d} regions during partitioning, thus confusing the classifiers, as several leaf nodes will refer to quite similar users or locations. A discussion of this point can be found in~\cite{han2012geolocation}.

\begin{figure}[t]
\centering
\includegraphics[width=7.5cm]{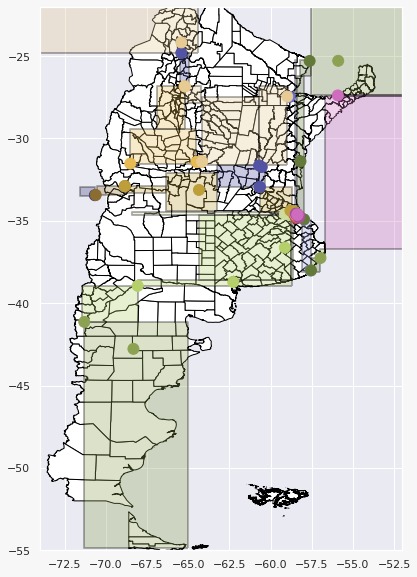}
\caption{\label{fig:kdtree}Division produced by a \textit{k-d} tree with at least 255 users per leaf node in our dataset \texttt{Twitter-ARG-Exact}.}
\end{figure}

For the \texttt{Twitter-ARG-BBox} dataset, we discarded using \textit{k-d} trees due to the former reason. For \texttt{Twitter-ARG-Exact}, instead, we evaluated both methods, but we found that the usage of \textit{k-d} trees was outperformed by city labels, due to the large number of samples in the Buenos Aires metropolitan area. In Figure~\ref{fig:kdtree} we show the division that a \textit{k-d} tree would produce on our dataset \texttt{Twitter-ARG-Exact}.

For \texttt{Twitter-US}, instead we will show results for both labels types, \textit{k-d} tree nodes and city names, for the sake of comparison.

\section{Models}

We propose 3 graph-based geolocation models: \textit{(a)} \texttt{RGCN-EXT}, a Relational Graph Convolutional Network that combines mentions and follower relationships as different layers of a multiplex,  \textit{(b)} \texttt{GraphSage-EXT}, an inductive learning model based on GraphSAGE and, \textit{(c)} \texttt{N2V-EXT}, a weighted graph embedding based on Node2vec+. All these models incorporate tweets' content information in some step of their pipelines, as described in the following subsections. We also incorporate two content-based methods for comparison: \texttt{BiLSTM-TXT}, based on a bidirectional Long Short-Term Memory (Bi-LSTM), and \texttt{Trans-TXT}, based on a transformer's encoder.

\subsection{\texttt{RGCN-EXT}}

Relational Graph Convolutional Networks (R-GCN's)~\citep{schlichtkrull2018modeling} constitute an extension to GCN's in which nodes can hold several types of relationships between them. In network theory, this concept is known as a \textit{multiplex} network.

Each neural layer in an R-GCN is computed as:
\[ h_i^{(l+1)} = \sigma \left( \sum_{r \in \mathcal{R}} \sum_{j \in \mathcal{N}_i^r} \frac{1}{c_{i,r}} W_r^{(l)}h_j^{(l)} + W_0^{(l)}h_i^{(l)} + b \right) \]

\noindent where $\mathcal{R}$ represents the set of relationship types and $\mathcal{N}_i^r$ represents the set of neighbors of node $i$ with relationship type $j$. We use the relationships obtained from the extended, weighted mention graph, and those of the extended, weighted follower graph when available.

The normalization constants $c_{i,r}$ can be learned, as well as the $W^{(l)}$ matrices and the bias, $b$.
The input vector $h_i^{(0)}$ for each node at the input layer is a feature vector for user $i$ denoting a score for each label. We use a ReLU activation function ($\sigma$) for the intermediate R-GCN layers and a softmax activation at the output layer.

The input scores $h_i^{(0)}$ are computed by training a transformer's encoder over the users' text view (as explained for the baseline models in Subsection~\ref{baselines}) and combining its predictions with those given by Location Indicative Words (LIW's). The list of LIW's in each training set is built using $\chi^2$ tests as proposed in~\cite{han2014text}.

\begin{figure*}[t]
\centering
\includegraphics[width=15cm]{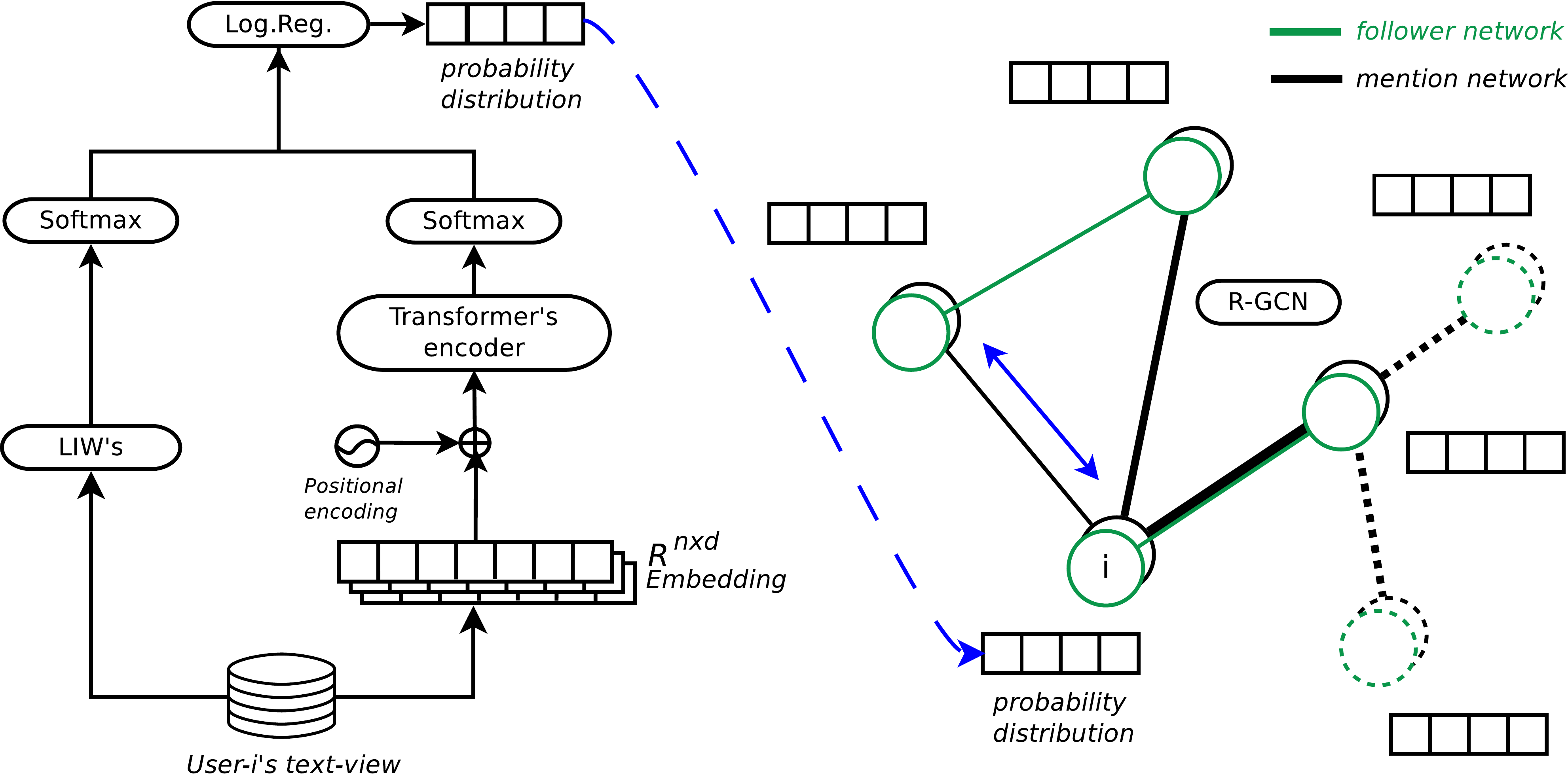}
\caption{\label{fig:rgcn}Components of the \texttt{R-GCN} model. On the left part of the diagram, the transformer's encoder classifier and Location Indicative Words (LIW's) are trained to output a probability for each discrete location. On the right, the multiplex network that combines the extended mentions (black) and extended followers (green) graphs, over which the four layers of the R-GCN are applied. The initial feature vectors for this stage are the probability distributions computed during the first stage.}
\end{figure*}

The full procedure is illustrated in Figure~\ref{fig:rgcn}; we implemented this model through the StellarGraph library~\cite{StellarGraph}, and we denote it as \texttt{RGCN-EXT} in our results.

\subsection{\texttt{GraphSAGE-EXT}}

GraphSAGE~\citep{hamilton2017inductive} is an inductive learning framework capable of generating embeddings based on each node's local neighborhood's features. During this process, the model learns a set of \textit{aggregator functions} that can later be used to make predictions for nodes that were not seen during training.

We apply this model to the extended, weighted mention graph using the MeanAggregator proposed in~\citep{hamilton2017inductive}.



The aggregator at each node is computed as:
\[
h_i^{(l+1)} = \sigma\left(W^{(l)}\cdot\left(h_i^{(l)}\Bigg\Vert\frac{\sum_{j\in \mathcal{N}_i}{h_j^{(l)}}}{|\mathcal{N}_i|}\right)\right).
\]

The input vectors $h_i^{(0)}$ are computed as in the \texttt{RGCN-EXT} model, i.e., by scores obtained from training a transformer's encoder over the users' text view combined with the predictions made by Location Indicative Words (LIW's).

In our \texttt{GraphSAGE-EXT} model, we use a ReLU nonlinearity as the $\sigma$ activation function, with a final softmax layer. We used the StellarGraph implementation~\cite{StellarGraph} of GraphSAGE.


\subsection{\texttt{N2V-EXT}}

Node2vec~\cite{grover2016node2vec} is a dimensionality reduction method that can learn node representations from graphs such that nodes at shorter distances in the graph tend to lie closer in the representation space; this is achieved by using biased random walkers.

In our model \texttt{N2V-EXT} we apply the weighted graph variant of Node2vec, Node2vec+~\cite{liu2021accurately}, to our extended mention graph and extended follower graph, using the implementation in~\cite{liu2021pecanpy}.
We combine the results of a logistic regression classifier for these embeddings with those of a transformer's encoder on the users' text view, by applying a logistic regression meta-classifier on the concatenation of both outputs.

\subsection{\label{baselines}Baseline content-based methods}

We also implement two baseline models that only consider the users' text view: \texttt{BiLSTM-TXT} and \texttt{Trans-TXT}.

In the \texttt{BiLSTM-TXT} model we use GloVe~\cite{pennington2014glove} to extract embeddings from the content and train a BiLSTM to predict the location probabilities.

In the \texttt{Trans-TXT} model we train a transformer's encoder~\cite{vaswani2017attention} over the content. The input of this model is formed by the $d$-dimensional GloVe embeddings of the content generated by the user; the embedding matrix is also updated during training. A positional encoding is added to the input embedding and the result, $H\in \mathbf{R}^{n\times d}$, enters the encoder, which applies a multi-head attention layer and obtains a single representation $o$ for each user:
\[
o = \text{MultiHead}(Q,H) = [\text{head}_1,...,\text{head}_h]W^O,
\]
\noindent where:
\[
\text{head}_i(Q, H) = \text{softmax}\left(\frac{QW_i^Q(HW_i^K)^T}{\sqrt{d_k}}\right)HW_i^V.
\]

The $W^Q_i$, $W^K_i$ and $W^V_i$ projections in each head $i$ are learnt during training, as well as the attention context $Q$ and the final projection $W^O$. The output of each head is of size $d_k = d/h$, and we use $h=10$ different heads for \texttt{Twitter-US} and $h=6$ for our dataset. The output of the attention layer passes through a feed-forward and a softmax layer. These results are combined with the predictions made by Location Indicative Words (LIW’s) to train a meta-classifier.

We chose these two types of methods as they had been previously used to achieve the highest results in the literature (see~\cite{huang-carley-2019-hierarchical}).

\section{Results}

In Table~\ref{tab:our_results} we show the performance of the three proposed models (\texttt{RGCN-EXT}, \texttt{GraphSAGE-EXT} and \texttt{N2V-EXT}) and the two baselines, for our introduced dataset. We report the \textit{Acc@100}, together with the mean and median errors. From these results we observe that the addition of network information produced a consistent improvement in performance, without any significant difference among the three methods. In this sense, we remark that \texttt{GraphSAGE-EXT} keeps the advantage of being an inductive algorithm, which makes it faster and more generalizable with respect to the other methods.

In Tables~\ref{tab:twitter_us_results} we show the results for the \texttt{Twitter-US} dataset, either using city names or \textit{k-d} tree nodes as labels. We differentiate between the performance metrics obtained under different types of labels in order to make a fair comparison. In this sense, the label choice can strongly affect the results, as we observed for our own dataset. We compare against other methods in the literature that either use content information, or combine it with the users' network. We do not compare against methods using metadata (e.g., the information in the user profile). Under tree-node labels, our results improved the state-of-the-art (\citep{miura2017unifying}), while for city labels we obtained a slightly smaller accuracy as compared to~\cite{huang-carley-2019-hierarchical}, who proposed a hierarchical classifier --first predicting county, then city-- that combined a Bi-LSTM on text embeddings and a network embedding.

\begin{table*}[t]
    \centering
    \begin{tabular}{llllclll}
    \toprule
    
     & \multicolumn{3}{c}{\texttt{Twitter-ARG-Exact}} & & \multicolumn{3}{c}{\texttt{Twitter-ARG-BBox}}\\
     \cline{2-4} \cline{6-8}
        Method & \small{Acc@100} $\uparrow$ & Mean $\downarrow$ & Median  $\downarrow$ & & \small{Acc@100} $\uparrow$ & Mean $\downarrow$ & Median $\downarrow$\\
    \midrule
    \midrule
        \textit{Content-based} \\
    \midrule
    \midrule
          \texttt{BiLSTM-TXT} & $0.750$ & $547.5$ & $5.2$ & & $0.496$ & $989.9$ & $191.6$ \\
     \texttt{Trans-TXT} & $0.764$ & $522.5$ & $4.9$ & & $0.532$ & $914.7$ & $51.6$ \\
    \midrule
    \midrule
        \textit{(Graph+content)-based} \\
    \midrule
    \midrule
    \texttt{RGCN-EXT} & $0.829$ & $\mathbf{367.5}$ & $\mathbf{3.8}$ & & $0.644$ & $\mathbf{690.6}$ & $6.4$ \\
     \texttt{GraphSAGE-EXT} & $0.823$ & $369.1$ & $\mathbf{3.8}$ & & $0.578$ & $798.4$ & $23.2$ \\
     \texttt{N2V-EXT} & $\mathbf{0.832}$ & $371.4$ & $3.9$ & & $\mathbf{0.653}$ & $719.2$ & $\mathbf{0}$ \\
    \bottomrule
    \end{tabular}
    \caption{
Cross-validated prediction results for the two Latin American datasets proposed here. We evaluate the results for our 3 proposed models combining network and content information, and we compare them with 2 content-based models.}
    \label{tab:our_results}
\end{table*}

\begin{table*}[t]
    \centering
    \begin{tabular}{llllllll}
    \toprule
    
     & \multicolumn{3}{c}{\texttt{Twitter-US} (city labels)} & & \multicolumn{3}{c}{\texttt{Twitter-US} (node labels)}\\
     \cline{2-4} \cline{6-8}
        Method & \small{Acc@100} $\uparrow$ & Mean $\downarrow$ & Median  $\downarrow$ & & \small{Acc@100} $\uparrow$ & Mean $\downarrow$ & Median $\downarrow$\\
    \midrule
    \midrule
    \textit{Content-based} \\
    \midrule
    \midrule
     \texttt{BiLSTM-TXT} & $0.508$ & $638.9$ & $150.8$ & & $0.500$ & $603.1$ & $161.7$ \\
     \texttt{Trans-TXT} & $0.523$ & $653.6$ & $132.4$ & & $0.513$ & $589.3$ & $146.7$ \\
    \citep{rahimi-etal-2017-neural}-\texttt{MLP} & $-$ & $-$ & $-$ & & $\mathbf{0.54}$ & $\mathbf{554}$ & $\mathbf{120}$ \\
    \citep{miura2017unifying}-\texttt{LR} & $-$ & $-$ & $-$ & & $0.527$ & $666.6$ & $121.1$ \\
    \cite{huang-carley-2019-hierarchical}-\texttt{HLPNN-Text} & $\mathbf{0.571}$ & $\mathbf{516.6}$ & $\mathbf{89.9}$ & & $-$ & $-$ & $-$ \\
    \midrule
    \midrule
    \textit{(Graph+content)-based} \\
    \midrule
    \midrule
     \texttt{RGCN-EXT} & $0.666$ & $408.7$ & $43.3$ & & $\mathbf{0.670}$ & $\mathbf{384.0}$ & $57.2$ \\
     \texttt{GraphSAGE-EXT} & $0.643$ & $432.1$ & $50.3$ & & $0.632$ & $424.3$ & $70.2$ \\
     \texttt{N2V-EXT} & $0.690$ & $421.0$ & $35.3$ & & $\mathbf{0.670}$ & $394.8$ & $57.0$ \\
    \citep{rahimi-etal-2017-neural}-\texttt{MADCEL-W-MLP} & $-$ & $-$ & $-$ & & $0.61$ & $515$ & $77$ \\
    \citep{miura2017unifying}-\texttt{MADCEL-B-LR} & $-$ & $-$ & $-$ & & $0.601$ & $582.8$ & $66.5$ \\
        \citep{miura2017unifying}-\texttt{SUB-NN-UNET} & $0.615$ & $481.5$ & $65$ & & $-$ & $-$ & $-$ \\
    
    \citep{P18-1187}-\texttt{MLP-TXT+NET}  & $-$ & $-$ & $-$ & & $0.66$ & $420$ & $\mathbf{56}$ \\
    \cite{huang-carley-2019-hierarchical}-\texttt{HLPNN-NET} & $\mathbf{0.708}$ & $\mathbf{361.5}$ & $\mathbf{31.6}$ & & $-$ & $-$ & $-$ \\
    
    \bottomrule
    \end{tabular}
    \caption{
Cross-validated prediction results for the \texttt{Twitter-US} dataset, either with GeoNames cities (left) or nodes of a \textit{k-d} tree (right) as prediction targets. When using node labels we consider the median of each region as its representative location, following~\citep{rahimi-etal-2015-twitter}. We evaluated the results for the 3 graph-based models and compared them against 2 proposed baseline models and others in the literature.}
    \label{tab:twitter_us_results}
\end{table*}

\section{\label{sec_rel_work}Related work}

One of the first works in the topic of user geolocation is the one by~\citep{cheng2010you}, which proposes a model based on the probabilities of using each word in different cities or states. After refining their model by identifying Location Indicative Words (LIW's) and smoothing the distributions, they reached an  score of $51\%$ of the users being assigned to a city closer than 100 miles to their ground truth location. This \textit{relaxed accuracy} measure was later adopted by the literature and referred to as \textit{Acc@100} (accuracy at less than 100 miles).

Due to the uneven geographical distribution of the population in many countries, predicting the exact location (city, county or state) makes the classification problem highly unbiased. ~\citep{roller2012supervised} was the first to propose using a grid based on a $k$-$d$ tree instead, in order to produce balanced classes. They obtained an \textit{Acc@100} of 0.346 on the \texttt{Twitter-US} dataset that they captured.

The selection of Location Indicative Words (LIW's) was later improved by~\citep{han2014text}, by comparing different statistical-based and  information theory-based methods. The authors tested their method on the \texttt{Twitter-US} and \texttt{Twitter-WORLD}~\cite{han2012geolocation} datasets, obtaining an \textit{Acc@100} of 0.450 and 0.270 respectively.

~\citep{chi2016geolocation} built off~\citep{han2014text}'s work by adding city mentions and the top 10k Twitter @mentions and \#hashtags to the LIW's. Using a multinomial Naive Bayes model, they obtained an accuracy of 0.225, on the \texttt{W-NUT} dataset. ~\citep{mahmud2014home} used an ensemble of statistical and heuristic classifiers, and followed a hierarchical approach to predict time zone, state, and then city. They captured a dataset of 1.5M tweets generated by 9.5k users in 100 cities, and obtained an accuracy of 0.58 at the city level.

Some later works enhanced the performance by adding some type of network information into the models, based on the location homophily principle: users who are closer are more likely to establish connections and interact. \citep{rahimi-etal-2015-twitter, rahimi-etal-2017-neural} were the first to incorporate the @mention network into the  prediction model: i.e., they built an undirected network in which two users $(u_1, u_2)$ are connected if one of them mentioned the other.
They discarded mentions of highly popular users (e.g., celebrities) assuming that they would not be predictive of the follower's location, and they added edges by transitivity when the intermediate nodes were not part of the training/test set. They used a label propagation model based on Modified Adsorption (MAD, ~\citep{talukdar2009new}), reaching an \textit{Acc@100} of 0.61 for \texttt{Twitter-US} and 0.53 for \texttt{Twitter-WORLD}, using a $k$-$d$ tree-based grid.

Other network-based models like~\citep{miura2017unifying} and~\citep{huang-carley-2019-hierarchical} achieved an \textit{Acc@100} of 0.70 and 0.72 on \texttt{Twitter-US} respectively, by using  neural network architectures and combining embeddings for the content, some metadata and the mention graph.
~\citep{P18-1187} also explored the usage of a Graph Convolutional Network (GCN,~\citep{kipf2017semi}) on the mention graph, obtaining an Acc@100 of 0.62 for \texttt{Twitter-US} y 0.54 for \texttt{Twitter-WORLD}.

Up to our knowledge, scarce works have used the follower network for location prediction in Twitter, possibly due to the fact that this type of data is difficult to obtain. ~\cite{li2012multiple} collected this information for $140k$ users obtaining a follower network with ~2 million edges; they fitted a probabilistic generative model obtaining an Acc@100 of $0.62$. ~\citep{rodrigues2016exploring} collected tweets from $12k$ users in Brazil, retrieving the last 200 tweets from each, and building a follower-followee graph with $125k$ edges. They obtained an accuracy of 0.652 when classifying the users into 10 Brazilian cities, using a Markov random field model that combined content and follower relationships. Finally, we mention that~\citep{hemamalini2018location}, proposed an implementation for a location service based on follower information, but did not analyze any performance metrics or datasets. A thorough survey of the location prediction problem in Twitter can be found in~\citep{zheng2018survey}.

\section{Conclusions}

In this work we proposed three graph-based methods for user geolocation that can propagate users' generated content through their extended mention and follower networks using different mechanisms: \texttt{RGCN-EXT}, \texttt{GraphSAGE-EXT} and \texttt{N2V-EXT}. In particular, the \texttt{GraphSAGE-EXT} model is trained under an inductive learning setting, thus avoiding retraining when new nodes appear in a real scenario.

We also proposed a new methodology for extending the mention graph with co-mentions towards third-users without overloading it, and we introduced the usage of the follower graph in some of our methods. Our results for the Latin America dataset suggest that the extended follower network, when available, can significantly improve the predictions as compared to the state-of-the art baseline models; i.e., both the \texttt{RGCN-EXT} and \texttt{N2V-EXT} models obtained the best results in this case.
The inclusion of the user text view in the models was performed through the encoder component of a transformer, this reduces the execution time as compared to other methods using recurrent neural network architectures. 

On the other hand, in the well-known \texttt{Twitter-US} dataset our results improved the state of the art when using \textit{k-d} tree nodes for building labels, while for city name levels, \texttt{N2V-EXT} stayed slightly below the results obtained by~\cite{huang-carley-2019-hierarchical} (significantly improving those obtained by~\citep{miura2017unifying}, though).

By using two different georeference fields in the Twitter API's --coordinates and bounding boxes-- we measured the plausibility of training location models using bounding boxes, which are much more available that the precise tweet coordinates. While previous datasets in the literature are based on coordinates, we built a version of our dataset based on bounding boxes and reached an Acc@100 accuracy of 0.65.
Also, we discussed the importance of analysing which is the best label construction technique in each case: for our proposed dataset, a large metropolitan area like Buenos Aires would be split into different nodes of a \texttt{k-d} tree. So, despite other advantages of the latter as the possibility of balancing class sizes, in this case it was more convenient to use cities as levels in order to obtain a smaller error. Also for the \texttt{Twitter-US} dataset, the best results were in general found at the city resolution level.

Future lines of research beyond this work include profiting of the multi-layer design of the R-GCN's in order to introduce new layers (e.g., separating mentions from comentions, or incorporating a retweet network) and modeling the propagation of partially available features (e.g., profile information) with techniques as the ones proposed by~\cite{rossi2021unreasonable}. Another interesting extension is to take into account users who are moving, either temporally (e.g., travelling) or due to migration or relocation. These factors could be taken into account by using embedding or diffusion methods based on temporal graphs.

\section*{Acknowledgements}

The authors acknowledge the financial support of a PICT 2019 grant from the \textit{Agencia Nacional de Promoción Científica y Tecnológica} (PICT 2019-01031).


\bibliographystyle{apalike}




\end{document}